\RequirePackage{lineno}
\documentclass[3p,twocolumn,UKenglish]{elsarticle}
\pdfoutput=1
\DeclareGraphicsExtensions{.pdf}

\biboptions{sort&compress}
\usepackage{atlaspackage}
\usepackage{xspace}
\usepackage{multirow}
\usepackage{amssymb}
\usepackage{amsmath}
\usepackage{lineno}
\usepackage{slashed}
\usepackage{graphicx}
\usepackage{amssymb}
\usepackage[preprint]{atlascover}
\usepackage{atlasphysics}
\usepackage{placeins}
\usepackage{array}
\usepackage[autostyle=true]{csquotes}
\newcommand*\patchAmsMathEnvironmentForLineno[1]{%
  \expandafter\let\csname old#1\expandafter\endcsname\csname #1\endcsname
  \expandafter\let\csname oldend#1\expandafter\endcsname\csname end#1\endcsname
  \renewenvironment{#1}%
     {\linenomath\csname old#1\endcsname}%
     {\csname oldend#1\endcsname\endlinenomath}}%
\newcommand*\patchBothAmsMathEnvironmentsForLineno[1]{%
  \patchAmsMathEnvironmentForLineno{#1}%
  \patchAmsMathEnvironmentForLineno{#1*}}%
\AtBeginDocument{%
\patchBothAmsMathEnvironmentsForLineno{equation}%
\patchBothAmsMathEnvironmentsForLineno{align}%
\patchBothAmsMathEnvironmentsForLineno{flalign}%
\patchBothAmsMathEnvironmentsForLineno{alignat}%
\patchBothAmsMathEnvironmentsForLineno{gather}%
\patchBothAmsMathEnvironmentsForLineno{multline}%
}
\usepackage{atlascontribute}

\newcommand{\papertype}{Letter}

\input{macros}

\hypersetup{colorlinks,breaklinks,pdftitle={ATLAS Draft Cover},pdfauthor={ATLAS Collaboration}}
\hypersetup{linkcolor=blue,citecolor=blue,filecolor=black,urlcolor=blue}

\AtlasTitle{
Measurement of the production of \nearby jets in lead-lead collisions at $\sqrtsnn = 2.76$~\TeV\ with the ATLAS detector
}

\AtlasAbstract{
This Letter presents measurements of correlated production of nearby jets in Pb+Pb collisions at $\sqrtsnn = 2.76$~TeV using the ATLAS detector at the Large Hadron 
Collider.
  The measurement was
performed using 0.14~${\rm nb}^{-1}$ of data recorded in 2011.
  The production of correlated jet pairs was quantified using the 
rate, \RdR, of ``\nearby'' jets that accompany ``\test'' jets within a given range of angular
distance, \dR, in the pseudorapidity--azimuthal angle plane. The jets were measured in the ATLAS calorimeter and were reconstructed using the \antikt\ algorithm with radius parameters \RTwo, $0.3$, and 
$0.4$. \RdR\ was measured in different \PbPb\ collision centrality bins,
characterized by the total transverse energy
measured in the forward calorimeters. A
centrality dependence of \RdR\ is observed for all three jet radii
with \RdR\ found to be lower in central collisions than in peripheral collisions. The ratios formed by the \RdR\ values in different centrality bins and the values in the 40--80\% centrality bin are presented.

}

\AtlasVersion{3.0}
\AtlasJournal{Physics Letters B}
\PreprintIdNumber{CERN-PH-EP-2015-143}

\begin{document}
\title{
\center{
Measurement of the production of \nearby jets in lead--lead collisions at $\sqrtsnn = 2.76$~\TeV\ with the ATLAS detector 
}}

\author{The ATLAS Collaboration}

\begin{abstract}
This Letter presents measurements of correlated production of nearby jets in Pb+Pb collisions at $\sqrtsnn = 2.76$~TeV using the ATLAS detector at the Large Hadron 
Collider.
  The measurement was
performed using 0.14~${\rm nb}^{-1}$ of data recorded in 2011.
  The production of correlated jet pairs was quantified using the 
rate, \RdR, of ``\nearby'' jets that accompany ``\test'' jets within a given range of angular
distance, \dR, in the pseudorapidity--azimuthal angle plane. The jets were measured in the ATLAS calorimeter and were reconstructed using the \antikt\ algorithm with radius parameters \RTwo, $0.3$, and 
$0.4$. \RdR\ was measured in different \PbPb\ collision centrality bins,
characterized by the total transverse energy
measured in the forward calorimeters. A
centrality dependence of \RdR\ is observed for all three jet radii
with \RdR\ found to be lower in central collisions than in peripheral collisions. The ratios formed by the \RdR\ values in different centrality bins and the values in the 40--80\% centrality bin are presented.

\end{abstract}
\maketitle 

\newlength{\fighalfwidth}
\setlength{\fighalfwidth}{0.49\textwidth}

\section{Introduction}
\label{sec:intro}
Experimental studies of jet production in \PbPb\ collisions at the LHC
can directly reveal the properties of the quark--gluon plasma
created in the collisions. One predicted consequence of quark--gluon plasma formation is ``jet quenching''
  that refers to the modification of parton showers initiated by hard-scattering processes 
which take place in the quark--gluon plasma~\cite{Majumder:2010qh}.
 Measurements of jet pairs at the LHC provided the first
direct evidence of jet quenching~\cite{Aad:2010bu, Chatrchyan:1327643}. In those measurements, the enhancement of transverse momentum imbalance of dijets in central \PbPb\ collisions was observed.
Measurements at the LHC of inclusive jet suppression \cite{jetsupp,Aad:2014bxa} and the variation of
the suppression with jet azimuthal angle with respect to the elliptic
flow plane \cite{Aad:2013sla} have shown that the transverse energy of jets is significantly
degraded and that the energy loss depends on the path length of the
parton shower in the plasma. 
  These dijet and single-jet measurements provide complementary information about the jet quenching process.
  The single jet measurements are sensitive to the average partonic energy-loss while the dijet measurements probe 
differences in the quenching between the two parton showers traversing the medium. Those differences can 
arise from the unequal path lengths of the showers in the medium or from fluctuations in the energy 
loss process itself.

   To help disentangle the contributions of these factors to the observed dijet asymmetries, the measurement of the correlations between jets that are at small
relative angles was performed.
\Nearby jet pairs include jets originating from the same hard
interaction, but also jets from different hard interactions. The
latter are not of interest in this analysis, and are subtracted
statistically. The remaining neighbouring jet pairs result primarily from hard radiation by the parton that occurs early in the process of the shower formation.   
Generally, two \nearby jets originating from the same hard scattering should have more similar path lengths in the medium compared to the two jets in the previous dijet measurement. 
  Therefore measuring \nearby jets could probe differences in their quenching that do not result primarily from difference in path length. More generally, measurements of the correlated production of jets in the same parton shower may provide more 
detailed insight into the modification of the parton shower in the quark--gluon plasma beyond the subsequent 
quenching of the resulting jets.

  This \papertype\ presents measurements of the production rate of \nearby jets in \PbPb\ collisions at $\sqrtsnn = 2.76$~\TeV\ characterized by the quantity \RdR\ introduced in Ref.~\cite{Abazov:2012lua}. The \RdR\ variable quantifies the rate of \nearby jets that accompany ``\test'' jets within a given range of angular distance, \dR, in the pseudorapidity--azimuthal angle ($\eta$--$\phi$)~plane\footnote{
    ATLAS uses a right-handed coordinate system with its origin at the nominal interaction point (IP) in the 
  centre of the detector and the $z$-axis along the beam pipe. The $x$-axis points from the IP to the centre 
  of the LHC ring, and the $y$ axis points upward. Cylindrical coordinates $(r,\phi)$ are used in the 
  transverse plane, $\phi$ being the azimuthal angle around the beam pipe. The pseudorapidity is defined in 
  terms of the polar angle $\theta$ as $\eta=-\ln\tan(\theta/2)$.}, where $\dR = \sqrt{(\Delta\eta)^2 + (\Delta\phi)^2}$. 
  Jets were reconstructed with the \antikt\ \cite{Cacciari:2008gp} algorithm using 
radius parameter values \RTwo, $0.3$, and $0.4$. 
  In events with  \test jets with transverse energy $\Et>70$~GeV, further jets are searched for within 
a certain angular distance from the \test jet.

The rate of the \nearby jets that accompany a \test jet, \RdR , is defined as
\begin{equation}
  \RdR(\ETtest,\ETnbj)  = \frac{\sum^{\Ntest}_{i=1} \Nnbji(\ETtest, \ETnbj, \dR)}{\Ntest (\ETtest)},
  \label{eq:1} 
  \end{equation}
  where \ETtest\ and \ETnbj\ 
are the transverse energies of the \test and \nearby jet, respectively; \Ntest\ is the number of \test jets in a 
given \ETtest\ bin and \Nnbj\ is the number of \nearby jets. Further, the \RdR\ quantity was used to define per-test-jet normalized spectra of neighbouring jets as
\begin{equation}
  \frac{\mathrm{d} \RdR}{\mathrm{d} \ETnbj} = \frac{1}{\Ntest} \sum_{i=1}^{\Ntest} \frac{\mathrm{d} \Nnbji}{ \mathrm{d} \ETnbj} (\ETtest, \ETnbj, \dR).
\label{eq:2}  
\end{equation} 

Previous measurements of the correlated production of \nearby jets were
performed by the D$\emptyset$ experiment in $p\bar{p}$ collisions at the
Tevatron \cite{Abazov:2012lua}. The measurements by D$\emptyset$ were intended to measure the strong
coupling constant, $\alpha_\mathrm{s}$, and to test its running over a large
range of momentum transfers. The measurements presented in this
\papertype\ use similar techniques and follow notations introduced in that measurement.

\section{Experimental setup}
\label{sec:experiment}

  The measurements presented in this \papertype\ were performed using the ATLAS inner detector, calorimeter, trigger and data acquisition systems~\cite{Aad:2008zzm}. The inner detector \cite{Collaboration:2010knc} measures charged particles within the 
  interval $|\eta|<2.5$. The inner detector is composed of silicon pixel detectors in the innermost layers, 
  followed by silicon microstrip detectors and a straw-tube tracker, all immersed in a 2~T axial magnetic 
  field provided by a solenoid. 
  The calorimeter system consists of a high-granularity liquid argon (LAr) electromagnetic (EM) calorimeter covering $|\eta|<3.2$, a steel/scintillator sampling hadronic calorimeter covering $|\eta| < 1.7$, a LAr hadronic calorimeter covering $1.5 <|\eta| < 3.2$. The hadronic calorimeter has three sampling layers longitudinal in shower depth and has a $\Delta \eta \times \Delta \phi$ granularity of $0.1 \times 0.1$ for $|\eta| < 2.5$ and $0.2 \times 0.2$ for $2.5 < |\eta| < 4.9$.\footnote{An exception is the third sampling layer that has a segmentation of $0.2 \times 0.1$ up to 
$|\eta| = 1.4$.} The EM calorimeters are segmented into three shower-depth compartments with an 
additional pre-sampler layer. The forward regions are instrumented with copper/LAr and tungsten/LAr forward calorimeters (FCal) covering $3.2 < |\eta| < 4.9$, optimised
for electromagnetic and hadronic energy measurements, respectively. Two minimum-bias trigger scintillators (MBTS) counters are located on each side at 3.56 m along the beamline from the centre of the ATLAS detector. The MBTS detect charged particles in the range $2.1 < |\eta| < 3.9$. Each MBTS counter is divided into 16 sections, each of which provides measurements of both the pulse heights and arrival times of energy deposits. The zero-degree calorimeters (ZDCs) are located symmetrically at $z = \pm 140$~m and cover $|\eta| > 8.3$. In \PbPb\ collisions the ZDCs measure primarily ``spectator'' neutrons, which originate from one of the incident nuclei and do not interact hadronically with nucleons of the other nucleus.

Minimum-bias \PbPb\ collisions were required either to have the transverse energy in the whole calorimeter exceeding 50 GeV at the Level-1 trigger or to have a track reconstructed in the inner detector in coincidence with ZDC signals on both sides.

  Events with high-\pt\ jets were selected using a combination of a minimum-bias Level-1 trigger and High Level Trigger (HLT) jet triggers. The Level-1 minimum-bias trigger required a total transverse energy measured in the calorimeter to be larger than 10~\GeV. 
The HLT jet trigger used the offline Pb+Pb jet reconstruction described in Section~\ref{sec:jetrec}, except for the application of the final hadronic energy scale correction. 
The HLT jet trigger selected events containing a \RTwo\ jet with $\ET > 20$~\GeV.

\section{Event selection and data sets}
\label{sec:analysis}

   This analysis used data from \PbPb\ collisions at a nucleon--nucleon centre-of-mass
   energy of $\sqrtsnn = 2.76$~TeV recorded by ATLAS in 2011. 
It utilizes data samples corresponding to a total integrated luminosity of $0.14~\invnb$. The minimum-bias sample was recorded with different prescales depending on the instantaneous luminosity in the LHC fill. The prescale indicates which fraction of events that passed the trigger selection were selected for recording by the DAQ. The minimum-bias trigger recorded an effective luminosity of $7~\invmub$. Events selected by the minimum-bias trigger and the jet triggers were required to have a reconstructed primary vertex with at least three associated tracks each with $\pt>500$~MeV and a difference between time of pulses from the two sides of the MBTS detector of less than 7~ns. 
A total of 51 (14.2) million minimum-bias triggered (jet-triggered) events passed the applied event selections and were used in the analysis.

 In heavy ion collisions, ``centrality'' reflects the overlap volume of the two colliding nuclei, controlled by the impact parameter of the collisions. The centrality of \PbPb\ collisions was characterized by \ETfcal, the total transverse energy measured in the FCal \cite{ATLAS:2011ah}. The centrality intervals were defined according to successive percentiles of the \ETfcal\ distribution ordered from the most central (highest \ETfcal) to the most peripheral collisions. Production of \nearby jets was
  measured in four centrality bins: 0--10\%, 10--20\%, 20--40\%, and 40--80\%, with the 40--80\% bin serving as the reference. The percentiles were defined after correcting the \ETfcal\ distribution for a 2\% 
minimum-bias trigger inefficiency that affects the most peripheral events, 
which are not included in this analysis.

The performance of the ATLAS detector and offline analysis in measuring jets in the environment of \PbPb\ 
collisions was evaluated using a large sample of Monte Carlo (MC) events obtained by overlaying simulated  PYTHIA \cite{Sjostrand:2006za} hard-scattering events 
onto randomly selected minimum-bias \PbPb\ events, recorded by ATLAS during the same data-taking period as the data used in this analysis.
PYTHIA version 6.423 with the AUET2B 
tune~\cite{ATL-PHYS-PUB-2011-009} was used. Three million PYTHIA events were produced for each of five 
intervals of the transverse momentum of outgoing partons in the $2\rightarrow 2$ hard-scattering 
process, with boundaries $17, 35, 70, 140, 280,$ and 560~\GeV. The detector response to the PYTHIA events 
was simulated using GEANT4~\cite{Agostinelli:2002hh,Aad:2010ah}, and the simulated hits were combined with the data 
from the minimum-bias \PbPb\ events before performing the reconstruction.

\section{Jet reconstruction and \nearby\ jet selection}
\label{sec:jetrec}

  Jets were reconstructed within the pseudorapidity interval $|\eta| < 2.8$.
  The jet reconstruction techniques described in Ref.~\cite{jetsupp} were used, 
and are briefly summarized here. The \antikt\ algorithm was first run in four-vector recombination mode, 
on $\Delta \eta \times \Delta \phi = 0.1\times 0.1$ logical towers. 
The energies in the towers were obtained by summing energies of calorimeter cells, calibrated at a scale set for electron showers, within the tower boundaries. Then, an iterative procedure was used to 
estimate a calorimeter layer- and $\eta$-dependent underlying event (UE) energy density, while excluding actual jets from 
that estimate.
  The UE energy was subtracted from each calorimeter cell within the towers included in the reconstructed 
jet. The subtraction accounted for a $\cos{2\phi}$ modulation in the UE energy density due to collective 
flow \cite{ATLAS:2011ah} of the medium using a measurement of the amplitude and phase of that modulation in the 
calorimeter.  The jet energies and momenta were calculated via a sum of all cells contained within the jets, treating each cell as a massless four-vector, using \ET\ values after the UE subtraction.
  A correction was applied to the reconstructed jet transverse energies to account 
for jets not excluded or only partially excluded from the UE estimate.
 The magnitude of the correction was typically a few percent but can be as large as 10\% for jets whose energies are fully included in the UE estimate. Then, a final $\eta$- and jet \ET-dependent hadronic energy scale calibration factor was 
applied~\cite{jetsupp}.
  
Separate from the calorimeter jets, ``track jets'' were reconstructed by applying the \antikt\ algorithm with \RFour\ to charged particles having $\pT > 4$ GeV. The track jets were used in conjunction with electromagnetic clusters to remove the contribution of ``UE jets'' generated by fluctuations in the underlying event. The technique is described in detail in Ref.~\cite{jetsupp}.

  In the MC simulation, the kinematics of the reference PYTHIA generator-level jets (hereafter called ``truth 
jets'') were reconstructed from PYTHIA final-state particles with the \antikt\ algorithm with radius
$d = 0.2, 0.3$, and 0.4 using the same techniques as applied in \pp\ analyses \cite{Aad:2011he}. PYTHIA truth jets were matched to the closest reconstructed jet of the same $d$ value within $\dR = 0.2$. The resulting matched jets were used to evaluate the jet energy resolution (JER), the
jet energy scale (JES), the jet angular resolution, and the jet reconstruction efficiency.

The \RdR\ measurement was performed with the sample triggered by the jet triggers. The measurement was done differentially in transverse energy of the test and \nearby jets, and in collision centrality. 
Five \ETtest\ intervals 70--80, 80--90, 90--110, 110--140, 140--300~GeV and four \ETnbj\ intervals, 30--45, 45--60, 60--80, 80--130~GeV were used.    
Furthermore configurations where all the \ET\ bins of the test jets or of the \nearby jets have the same upper bound of 300 GeV were also used in this analysis. The number of bins and their boundaries were chosen to minimize the impact of the limited number of events in the data while preserving the ability to infer the trends in the measured distributions. 
For each jet radius, neighbouring jets are considered
if they lie within a specific annulus in $\dR$ around
the test jet: 
  $0.5< \dR < 1.6$, $0.6< \dR < 1.6$, and $0.8< \dR < 1.6$ for 
\RTwo, \RThree, and \RFour\ jets, respectively. 
  The inner edge of each annulus was chosen to avoid possible overlap of test and neighbouring jets, and the outer edge value ($\lesssim \pi/2$) rejects \nearby jets in the hemisphere opposite to the test jet and maximizes the number of events.
Choosing a maximum $\dR$ of 1.6 restricts the pseudorapidity range of \test jets to 
$|\eta| < 1.2$, yielding approximately $87\times 10^{3}$ \RFour\ test jets with $\pt > 80$~\GeV\ analysed in 0--10\% central events and $37\times 10^{3}$ test jets in 40--80\% peripheral events.

To quantify the centrality dependence of the \nearby jet yields, the 
ratio $\RRdR\equiv\RdR|_{\mathrm{cent}}/\RdR|_{40-80}$ is calculated as the ratio of \RdR\ measured in each centrality bin to \RdR\ measured in the reference (40--80\%) bin.

\section{Corrections to \nearby jet rates}
\label{sec:perform}

  The raw rates of \nearby jets include a
   contribution from \nearby 
jets that originate from different hard partonic interactions in the same \PbPb\ 
collision.
  This combinatorial background is present both in the MC simulation and in the data and must be subtracted. 
It is largest in the low \ETnbj\ bins and it increases with increasing centrality of the collision, since the probability for the presence of two independent hard scatterings in one Pb+Pb collision is expected 
to increase with the number of binary collisions. 
  The combinatorial background is estimated using the differential yield of inclusive 
jets ($\mathrm{d}^{3}N_{\mathrm{jet}}/\mathrm{d}\eta \mathrm{d}\phi \mathrm{d}\Et$) 
evaluated in minimum-bias Pb+Pb events. 
To each event considered a weight is assigned such that the event sample obtained from the minimum-bias trigger has the same centrality distribution as the sample collected by the jet trigger. This estimated background needs to be corrected for a geometrical bias 
present in the case where the combinatorial jet overlaps with a real \nearby 
jet or when two combinatorial jets overlap. These biases were removed by applying a multiplicative correction factor to background distributions prior 
to the subtraction.	 
  This multiplicative factor was derived from the reconstruction efficiency of 
two \nearby jets evaluated as a function of their angular separation in the 
annulus. In that evaluation, one jet was required to originate from PYTHIA's 
hard scattering and the other jet was required to originate from the 
minimum-bias data in the overlay. 
The impact of this correction on the final subtracted distribution is smaller than 0.5\%. 

The combinatoric jet kinematics may also be affected by the presence of a test jet. To control this influence, a study comparing the combinatoric jets from the overlay MC events with the same jets in the original minimum-bias data was performed. This study resulted in an additional correction, independent of centrality and jet \ET, that decreases the combinatorial background by 1.5\%. 
The $\pm$1.2\% uncertainty on the correction originates from the limited number of events and was included in the systematic uncertainties.

In order to account for the effect of the azimuthal dependence of jet yields~\cite{Aad:2013sla}, the combinatorial background was reweighted to take into account the measured azimuthal distributions of test jets as well as combinatorial jets. 
The change of the raw subtracted distribution in central collisions and low \ETnbj\ bins after the reweighting is at the level of 1\% and decreases with increasing centrality of the collision and \ETnbj.

  The background is subtracted from raw \RdR\ distributions both in the data and in the 
MC simulation, allowing an evaluation of the effectiveness of the subtraction using the MC simulation. 
  The signal-to-background ratio strongly depends on the centrality of the collision and 
\ETnbj . In 0--10\% central collisions, the signal-to-background ratio can be as low as 0.15 for the most extreme case of $30 < \ETnbj < 45$~GeV, and increases to approximately 0.8 for $60 < \ETnbj < 80$~GeV.

  The raw subtracted \RdR\ distributions are affected by the jet energy resolution.
  The combination of the jet energy resolution and the steeply falling \ET\ spectrum produces a net migration of jets 
from lower \ET\ to higher \ET\ values such that a jet reconstructed with a given \ETrec\ corresponds, 
on average, to a lower truth jet \ET, \avgETtrue. The relationship between \avgETtrue\ and \ETrec\ was 
evaluated in simulated events for the different centrality bins and three jet radii used in the analysis. 
The extracted relationships were used to correct for the average shift 
in the \test jet energy. 
  No correction due to the jet reconstruction efficiency for the \test jets is needed, since the analysis 
operates in the transverse energy region where the jet reconstruction is fully efficient. No correction due 
to jet trigger efficiency is needed either since the plateau of the jet trigger efficiency is reached for 
all test jets, except for \RFour\ jets with $\ETtest < 90$~GeV in the 0--10\% and 10--20\% centrality bins. 
  In the region $70<\Et<90$~GeV, the jet trigger efficiency is above 85\%. 
A systematic uncertainty is applied to describe the effect of the lower jet reconstruction efficiency.
  
  The impact of the jet energy resolution, reconstruction efficiency, and angular resolution on \nearby jet yields is corrected for by applying bin-by-bin unfolding to the raw subtracted \RdR\ distributions. For each measured \RdR\ distribution, two corresponding MC distributions are evaluated, one using truth jets and the other using jets after the detector simulation. The ratio of these two MC distributions provides a correction factor which is then applied to the data. 

The bin-by-bin correction factors are derived from the MC simulation where the reconstructed jets were matched to the truth jets. To account for the impact of the jet angular resolution, the truth jet is required to lie within a given annulus while the reconstructed jet is allowed to fall outside of the annulus.

\begin{figure}
\centerline{
\includegraphics[width=\fighalfwidth]{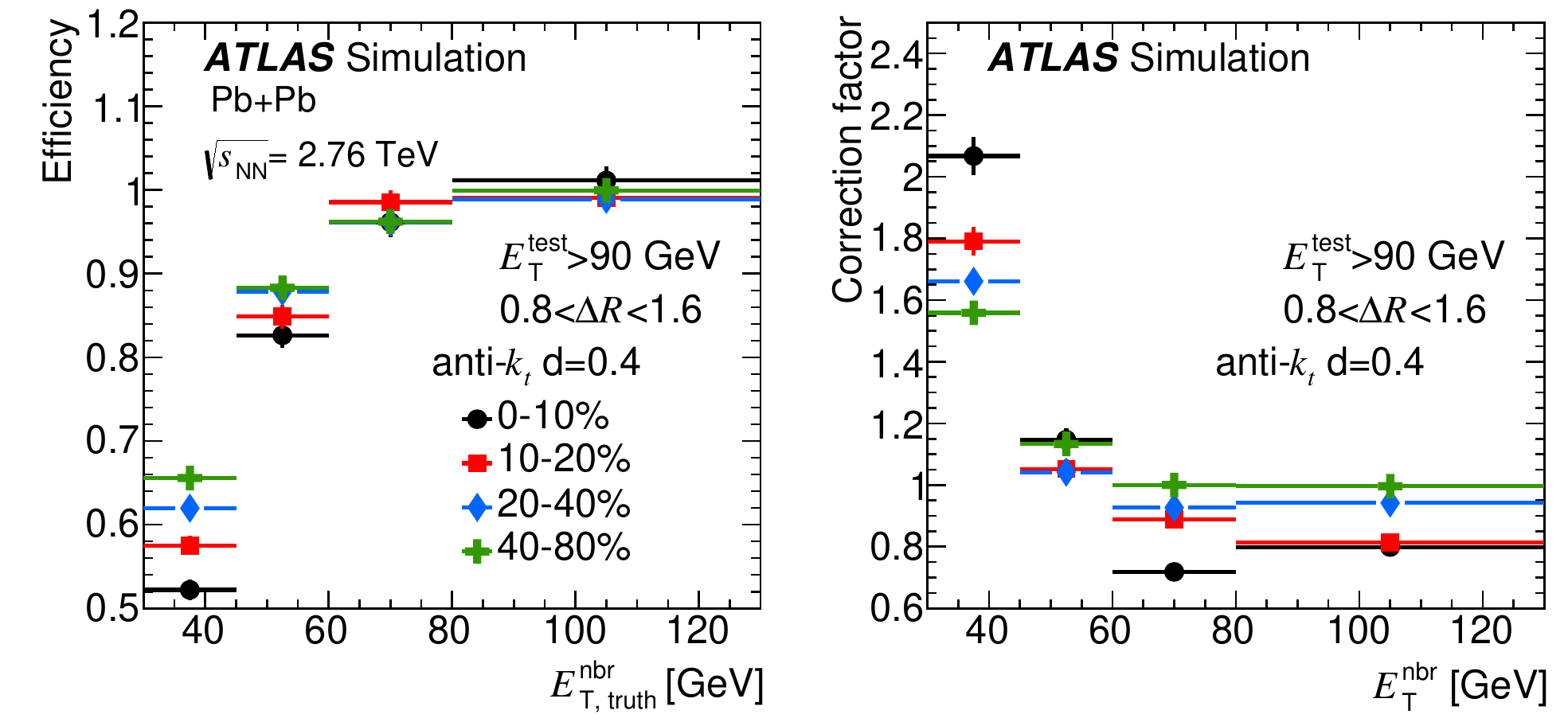}
}
\centerline{
\includegraphics[width=\fighalfwidth]{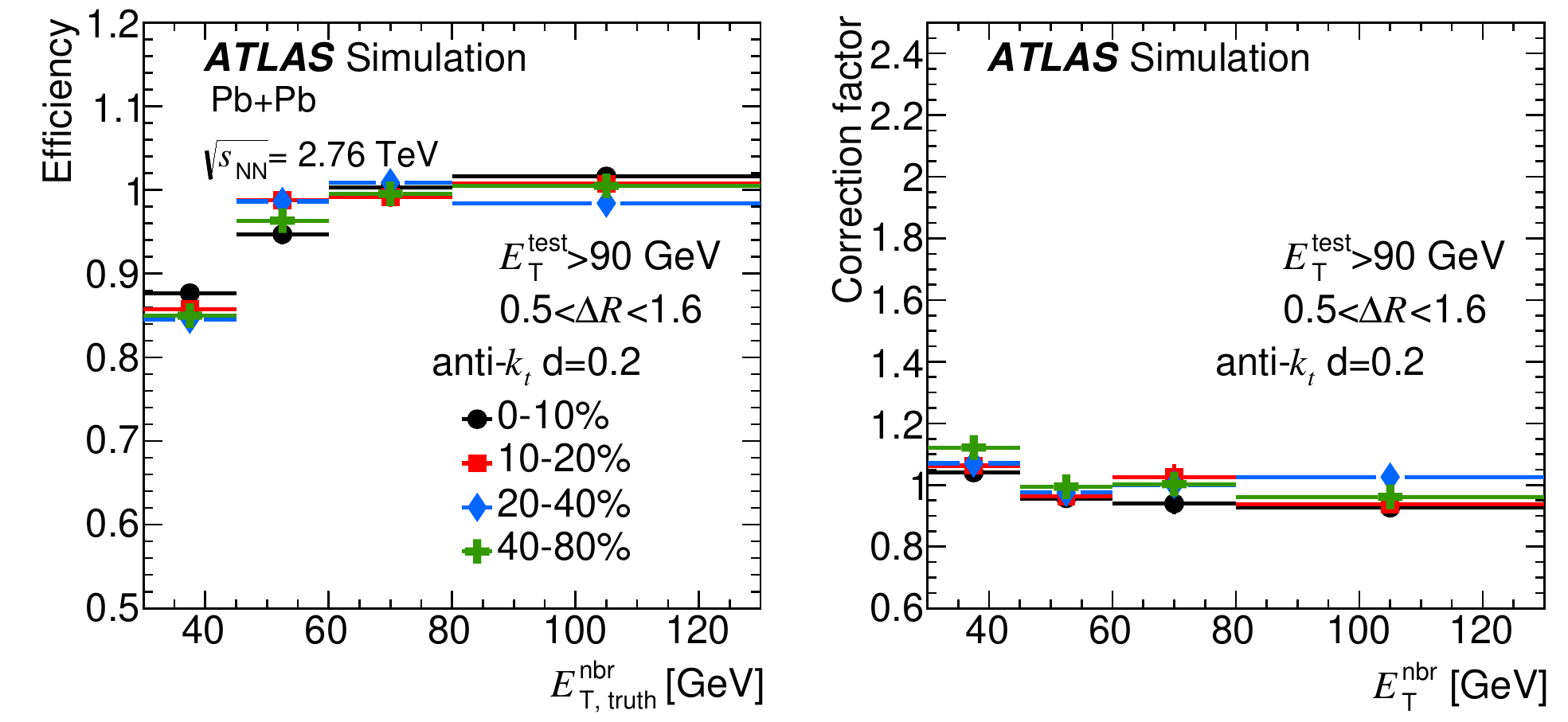}
}
\vspace{-0.15in}
\caption{
Reconstruction efficiency (left) of neighbouring jets as a function of the transverse energy of the neighbouring jet at Monte Carlo generator level \ETnbjtrue\ and bin-by-bin correction factors for the distributions of the neighbouring jet rate \RdR\ (right) as a function of the reconstructed transverse energy of the neighbouring jet \ETnbj. Plots for two different jet radii, \RFour\ (upper plots) and \RTwo\ (lower plots) are shown and the transverse energy of the test jet \ETtest\ is required to exceed 90 GeV. The four different centrality bins are denoted by different markers in each plot. The vertical error bars represent statistical uncertainties. 
}
\label{fig:performance}
\end{figure}

  Examples of jet reconstruction efficiencies for \nearby jets and the bin-by-bin correction factors are shown in Fig.~\ref{fig:performance} for different 
centrality selections and for two choices of jet radii: \RFour\ and \RTwo. 
Generally, the jet energy resolution in central (\mbox{0--10\%}) collisions for \RFour\ jets has comparable
contributions from UE fluctuations and the ``intrinsic'' resolution of the calorimetric jet
measurement. The fluctuations in the UE are approximately two times smaller for
\RTwo\ jets than they are for \RFour\ jets. Thus, the distributions measured using
\RTwo\ jets are far less sensitive to the effects of the jet energy resolution, and consequently the resulting
bin-by-bin correction factors for those distributions exhibit only a modest centrality dependence.
  The difference in the jet reconstruction efficiency between the 
two choices of jet radii is also significant -- the efficiency for \RTwo\ jets plateaus around 30 GeV, where it is still rising rapidly for \RFour\ jets.

The jet angular resolution is determined in MC simulation as the standard deviation of the difference in angles between truth and reconstructed jets. 
In both $\eta$ and $\phi$ it reaches 0.008 in 0--10\% collisions and 0.005 in the 40--80\% centrality bin for \RFour\ jets with \ET\ = 30 GeV. The angular resolution improves with increasing jet \ET\ and reaches 
  0.004 (resolution in $\eta$) and 0.002 (resolution in $\phi$) at $\ET=200$~GeV, independently of centrality.

\begin{figure}
\centerline{
\includegraphics[width=\fighalfwidth]{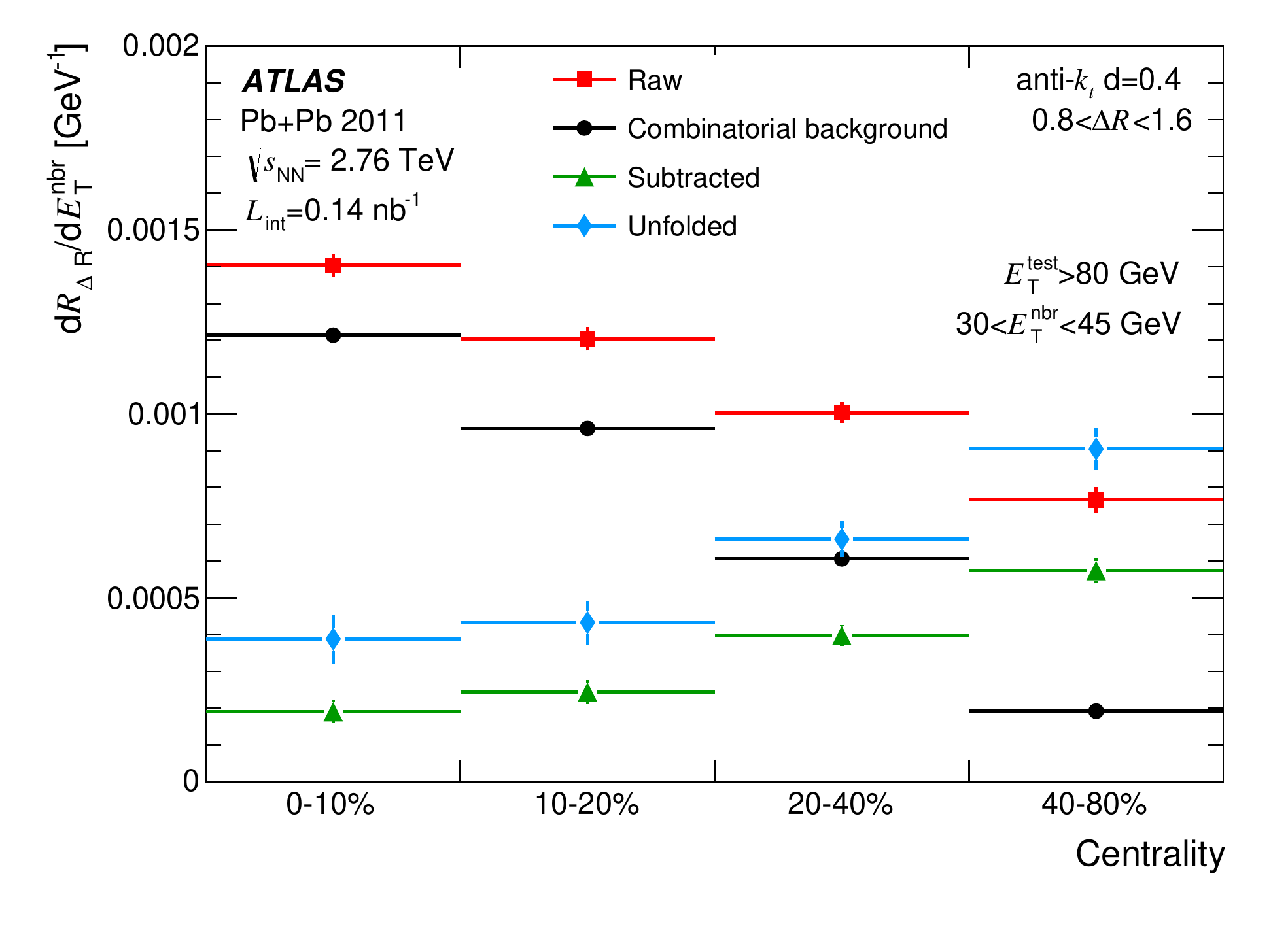}
}
\vspace{-0.15in}
\caption{
  Summary of how the corrections impact the ${\mathrm{d} \RdR}/{\mathrm{d} \ETnbj}$ distribution measured in the data in different centrality bins. The ${\mathrm{d} \RdR}/{\mathrm{d} \ETnbj}$is shown for \RFour\ jets for $\ETtest > 80$~GeV and in the interval $30 < \ETnbj < 45$~GeV. Squares show the raw ${\mathrm{d} \RdR}/{\mathrm{d} \ETnbj}$ prior to the UE subtraction, circles show the combinatorial background, triangles show the subtracted ${\mathrm{d} \RdR}/{\mathrm{d} \ETnbj}$ prior to unfolding by applying the bin-by-bin correction factors, and diamonds show the unfolded ${\mathrm{d} \RdR}/{\mathrm{d} \ETnbj}$. Vertical error bars on the combinatorial background, raw, and subtracted distributions represent statistical uncertainties. Vertical error bars on the unfolded distribution represent the combined statistical uncertainty from the unfolding and from subtracted distributions.
  }
\label{fig:CorrectionFlow}
\end{figure}

   The impact of these corrections on \RdR\ distributions measured in the data in different centrality bins is shown in Fig.~\ref{fig:CorrectionFlow}. The figure shows the raw \RdR\ distribution, and the combinatorial background, subtracted, and final unfolded \RdR\ distributions for \RFour\ jets with $\ETtest > 80$~GeV and in the lowest \ETnbj\ interval, $30 < \ETnbj < 45$~GeV. The raw \RdR\ distribution increases from peripheral to central collisions. The increase of the combinatorial background is steeper than the increase of the raw distribution. Therefore, a decrease of subtracted \RdR\ with increasing collision centrality is observed. The shape of the \RdR\ distribution remains unchanged when the bin-by-bin correction is applied.    

\section{Systematic uncertainties}
\label{sec:systerr}
Systematic uncertainties in the measurement of \RdR\ distributions and their ratios, \RRdR, arise from the uncertainty on the jet energy scale, jet energy resolution, angular resolution, bin-by-bin unfolding, centrality, combinatorial background and jet trigger efficiency. 
The impact of uncertainties on the jet energy scale, jet energy resolution and jet angular resolution was assessed by constructing new bin-by-bin correction factors with a systematically varied relationship between the reconstructed
and truth jet kinematics. The resulting uncertainties on \RdR\ and \RRdR\ were evaluated from their changed values obtained with modified jet energy scale, jet energy resolution and jet angular resolution dependencies.

The systematic uncertainty due to the jet energy scale is composed of two parts: an absolute, centrality-independent component, and a centrality-dependent component. The uncertainty on \RdR\ from the jet energy scale uncertainty is evaluated by shifting all reconstructed jet transverse energies by $\pm1$~standard deviation of the jet energy scale uncertainty. The absolute component is determined from the \textit{in situ} studies of the calorimeter
response; systematic variations of the jet response in the MC
simulation~\cite{Aad:2011he}; and from studies of the relative energy scale difference between the jet reconstruction procedure in heavy-ion collisions, and the procedure used for inclusive jet measurements in 2.76~TeV and 7~TeV \pp\ collisions~\cite{Aad:JESConf}. The magnitude of the uncertainty on the \RdR\ from the absolute jet energy scale uncertainty varies from 2\% to 15\% as a function of \Et\ and radius of the jet. 
The centrality-dependent component of jet energy scale uncertainty~\cite{Aad:2014bxa} was estimated using the PYQUEN MC generator~\cite{Lokhtin:2005px} to provide a sample of jets with modified fragmentation functions consistent with measurements in quenched jets performed by ATLAS \cite{Aad:2014wha} and CMS \cite{Chatrchyan:2012gw}.

The centrality-dependent jet energy scale uncertainty reaches 1\% for 0--10\% central collisions and less than 0.25\% for 40--80\% peripheral collisions.
  The uncertainty on \RdR\ originating from the centrality-dependent component of the jet energy scale 
uncertainty increases from less than 1\% in peripheral collisions to 3\% in central collisions.

The effect of the jet energy resolution uncertainty was evaluated by applying modified bin-by-bin correction factors where the reconstructed jet \Et\ was smeared. The uncertainty on the jet energy resolution is dominated by the uncertainty in the detector response. Thus, the procedure used for jet measurements in the 7~TeV \pp\ collisions~\cite{Aad:2011he} is used. 
  The smearing factor is evaluated using an \textit{in situ} technique
involving studies of dijet energy balance. The
systematic uncertainty on \RdR\ due to the jet energy resolution varies from 1\% to 4\%
depending on the jet \ET. 
  The centrality-independent jet energy scale uncertainty and the uncertainty from jet energy resolution tend to cancel
in the \RRdR\ ratios since both the numerator and denominator in the ratios are affected to a
similar degree by the variations accounting for the uncertainties.

The systematic uncertainty on combinatorial contributions originates from the previously noted uncertainty on the correction factor taking into account the difference between jets in minimum-bias events and combinatorial jets in the overlay. The resulting uncertainty reaches $\sim 8 \%$ in 0--10\% central collisions at low \ETnbj\ and rapidly 
decreases with decreasing centrality or increasing \ETnbj.

The systematic uncertainty associated with the bin-by-bin unfolding is connected with possible differences in the spectral shape between the data and the MC simulation. To achieve better correspondence with the data,
the simulated jet spectrum was reweighted to match the spectral shape in the data before deriving the bin-by-bin
correction factors as described above. Conservatively, the entire change in \RdR\ and \RRdR\ 
induced by the use of reweighted bin-by-bin correction factors is taken as a systematic uncertainty. Typically, this results in 1--2\% uncertainty on \RdR. A maximum uncertainty of 5\% is reached in 0--10\% 
central collisions for \RdR\ evaluated for \nearby jets with $\ETnbj > 30$~GeV.

The uncertainty on the centrality estimation originates from the uncertainty on the estimated inefficiency of the minimum-bias trigger. 
The analysis was repeated with modified centrality bins assuming 100\% minimum-bias trigger efficiency. 
The resulting uncertainty is typically smaller than 5\% with a mild \ET\ dependence and a negligible centrality dependence. 

The uncertainty associated with the jet angular resolution is estimated using modified bin-by-bin 
correction factors where the reconstructed jet $\eta$ and $\phi$ is smeared to reflect a 
up to $\sim$15\% centrality and \ET\ dependent uncertainty on the angular resolution.
  The uncertainty on the jet angular resolution was estimated by comparing the angular distance between track jets and the closest calorimetric jet in the data and in the MC simulation. The magnitude of uncertainty on \RdR\ from the jet angular resolution is smaller than 2\%.

  The systematic uncertainty on the jet trigger efficiency covers 
  a possible bias caused by selecting test jets in the region where the jet trigger is not fully 
efficient.
  This is the case for the \RFour\ jets with $\ETtest<90$~GeV reconstructed in the 0--10\% and 10--20\% 
centrality bins. For that \ET\ region, the systematic uncertainty was determined as the difference between the trigger efficiencies for inclusive jets and jets that were required to have a neighbouring jet. This trigger efficiency difference is less than 5\% and is independent of the \ETnbj.

\begin{figure}
\centerline{
\begin{tabular}{cc}
\includegraphics[width=\fighalfwidth]{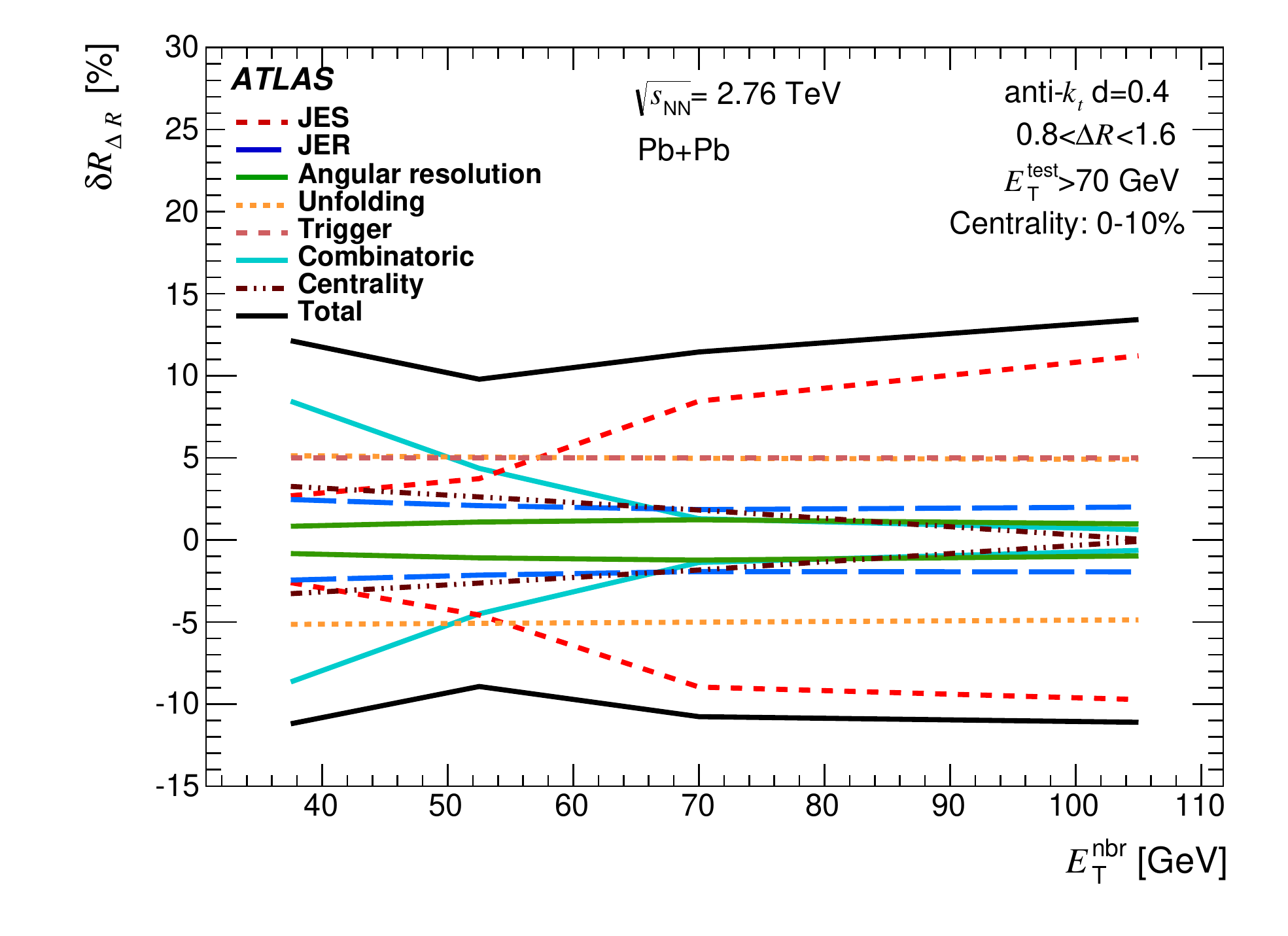}
\end{tabular}}
\vspace{-0.15in}
\caption{
  Summary of the relative systematic uncertainties, in \%, on the \RdR\ distributions ($\delta \RdR$). 
  The systematic uncertainties due to the jet energy scale (JES), the jet energy resolution (JER), the jet angular resolution, unfolding, jet trigger efficiency, combinatorial contributions and centrality are shown for \RFour\ jets with $\ETtest > 70$~GeV in 0--10\% central 
collisions.
  }
\label{fig:Systematics1}
\end{figure}

To avoid statistical fluctuations in the values of systematic uncertainties, the weak \ETnbj\ dependence of the uncertainties is smoothed by a second-order polynomial. Systematic uncertainties on \RdR\ for \RFour\ jets are summarized in Table~\ref{tbl:syst} for the 0--10\% and 40--80\% centrality bins. The table shows the maximum values of uncertainties for \RdR\ and for \RRdR. The total systematic uncertainties for jets with the other two jet radii are smaller than those shown in the table. For the 0--10\% centrality bin these systematic uncertainties are also plotted in Fig.~\ref{fig:Systematics1} as a function of \ETnbj.

\begin{table}[ht]
\centering
\begin{tabular}{|c||c|c|c|} \hline
                   & \multicolumn{2}{c|}{$\delta$\RdR }           & $\delta$\RRdR          \\ \cline{2-4}
                   & 0--10\%               & 40--80\%             & 0--10\%                \\ \hline
 JES               & 15\%                  &  11\%                 & 7 \%                    \\ \hline
 JER               & 4\%                   &  2\%                 & 2\%                    \\ \hline
 Angular           & \multirow{2}{*}{2\%}  & \multirow{2}{*}{0.5\%} & \multirow{2}{*}{2\%}   \\ 
 resolution        &                       &                      &                        \\ \hline
 Unfolding         & 5\%                   & 2\%                  & 5\%                    \\ \hline
 Centrality        & 6\%                   & 6\%                  & 6\%                    \\ \hline
 Combinatoric      & 8\%                   & $<$0.5\%                   & 8\%                    \\ \hline
 Trigger           & 5\%                   & --                   & 5\%                    \\ \hline
\end{tabular}
  \caption{
  Maximum systematic uncertainties on \RdR\ ($\delta$\RdR) and on the ratio of \RdR\ in central collisions and in peripheral (40--80\%) collisions \RRdR\ ($\delta$\RRdR) for \RFour\ jets in the 
0--10\% and 40--80\% centrality bins. The systematic uncertainty on the trigger is applicable only for $\ETtest < 90$~GeV.}
  \label{tbl:syst}
\end{table}

\begin{figure*}[th]
\centerline{
\includegraphics[width=0.9\textwidth]{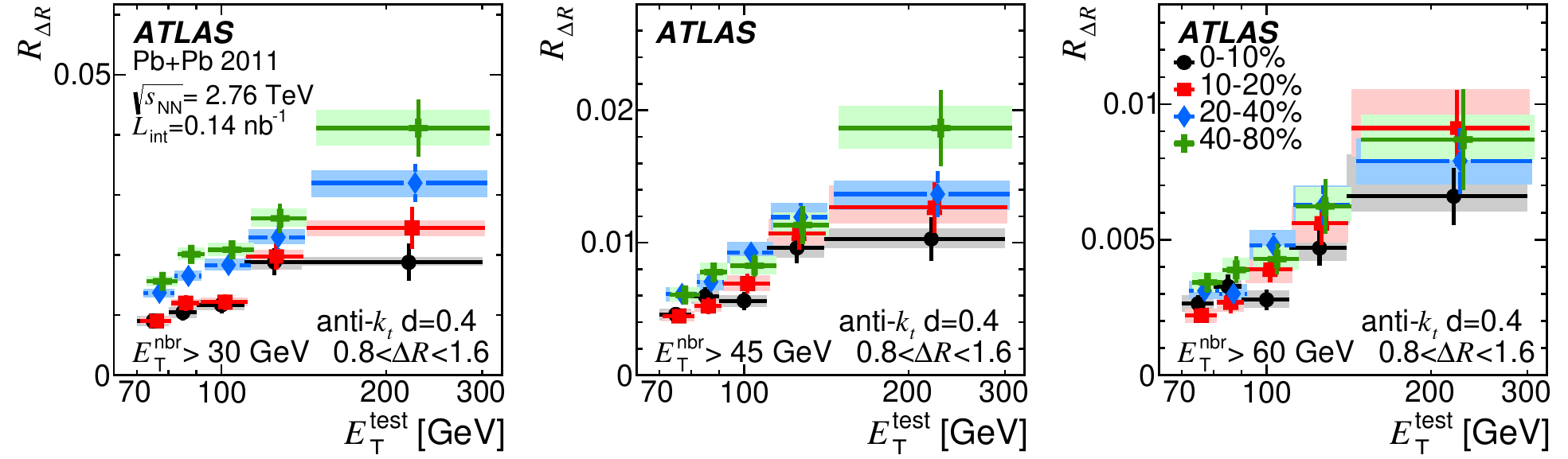}
}
\centerline{
\includegraphics[width=0.9\textwidth]{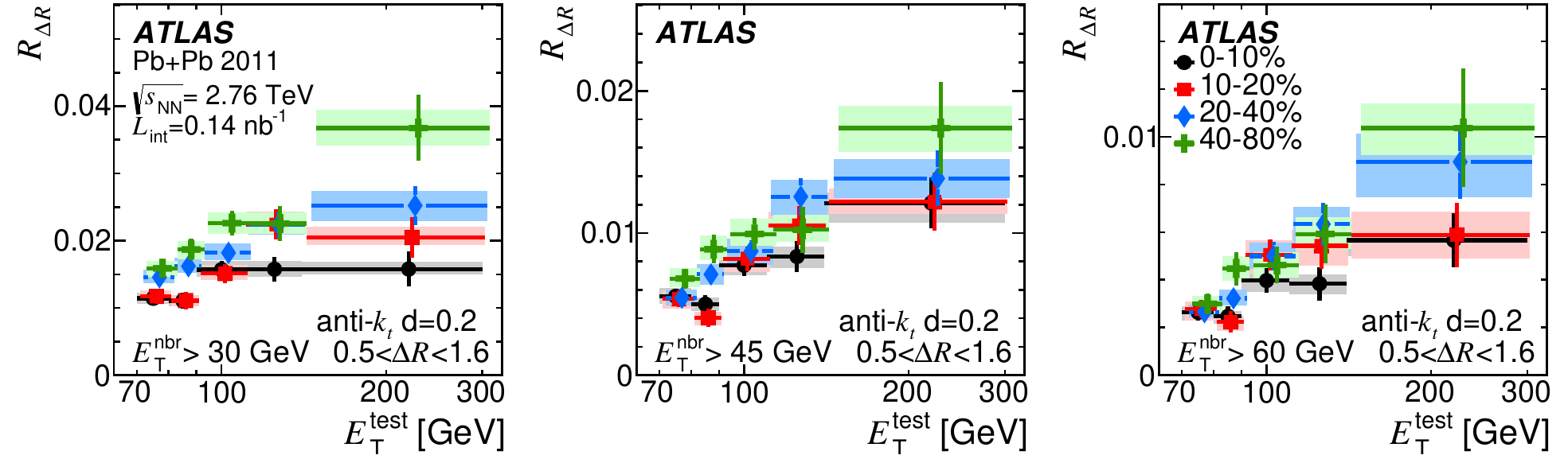}
}
\vspace{-0.15in}
\caption{ 
  \RdR\ distributions for \RFour\ jets (upper) and \RTwo\ jets (lower) evaluated as a function of 
\ETtest . The three different columns show \RdR\ distributions evaluated for three different lower bounds on the \nearby-jet transverse energy, $\ETnbj > 30$, 45, and 60~\GeV. The four 
different centrality bins are denoted by different markers in each plot. The shaded bands 
indicate systematic uncertainties, vertical error bars represent statistical 
uncertainties. 
  The data points and horizontal uncertainties for the 10--20\%, 20--40\%, and 40--80\% centrality bins are shifted along the
horizontal axis with respect to the 0--10\% centrality bin for clarity.
  }
\label{fig:OJR4R2}
  \end{figure*}

  \begin{table}
  \centering
  \begin{tabular}{| l || c |} \hline
               & 0--10\%         \\ \hline 

 $d=0.4$   & $2.73 \pm 0.23(stat.) \pm 0.12(syst.)$    \\ \hline
 $d=0.3$   & $2.83 \pm 0.16(stat.) \pm 0.14(syst.)$    \\ \hline
 $d=0.2$   & $2.81 \pm 0.15(stat.) \pm 0.15(syst.)$    \\ \hline\hline
 
                  &  10--20\%        \\ \hline 
 
  $d=0.4$   & $2.85 \pm 0.17(stat.) \pm 0.13(syst.)$    \\ \hline
  $d=0.3$   & $2.51 \pm 0.15(stat.) \pm 0.11(syst.)$    \\ \hline
  $d=0.2$   & $2.56 \pm 0.16(stat.) \pm 0.12(syst.)$    \\ \hline\hline
 
 &  20--40\%  \\ \hline
 $d=0.4$   & $2.90 \pm 0.12(stat.) \pm 0.10(syst.)$    \\ \hline
 $d=0.3$   & $2.91 \pm 0.11(stat.) \pm 0.09(syst.)$    \\ \hline
 $d=0.2$   & $2.62 \pm 0.13(stat.) \pm 0.10(syst.)$    \\ \hline
 
  &  40--80\% \\ \hline
  $d=0.4$    & $3.26 \pm 0.15(stat.) \pm 0.13(syst.)$   \\ \hline
  $d=0.3$    & $3.24 \pm 0.15(stat.) \pm 0.11(syst.)$   \\ \hline
  $d=0.2$    & $2.99 \pm 0.17(stat.) \pm 0.11(syst.)$   \\ \hline
  
  \end{tabular}
  \caption{
    Power-law index $n$ extracted from fits of ${\mathrm{d} \RdR}/{\mathrm{d} \ETnbj}$ distributions to a power-law function \\ $\propto1/\Et^{n}$ for $\ETtest  > 90$~\GeV, for four bins in centrality and three jet radii. The resulting fit error takes into account 
 combined statistical and systematic uncertainties.
  }
  \label{tbl:index}
  \end{table}

\section{Results}
\label{sec:results}

The \RdR\ measurement is performed differentially in collision centrality, transverse energy of the \test 
jet, \ETtest , and transverse energy of the \nearby jet, \ETnbj, as described Sec.~\ref{sec:jetrec}. The measured distributions are divided into four centrality bins, 0--10\%, 10--20\%, 20--40\%, and 40--80\%. Fig.~\ref{fig:OJR4R2} shows the fully corrected \RdR\ distributions for \RFour\ and \RTwo\ jets 
evaluated as a function of \ETtest . The distributions are shown for four centrality bins and three different lower bounds on the \nearby-jet transverse energy, $\ETnbj > 30$, 45, and 60~\GeV. The shaded error 
bands on the plots indicate the systematic uncertainties discussed in Section~\ref{sec:systerr}. The \RdR\ 
distribution exhibits an increase with increasing \ETtest. Sizeable differences between the four different centrality bins are observed for all three jet radii. The yield of \nearby jets is suppressed as the centrality of the collision increases.

\begin{figure*} [th]
\centerline{
\includegraphics[width=0.8\textwidth]{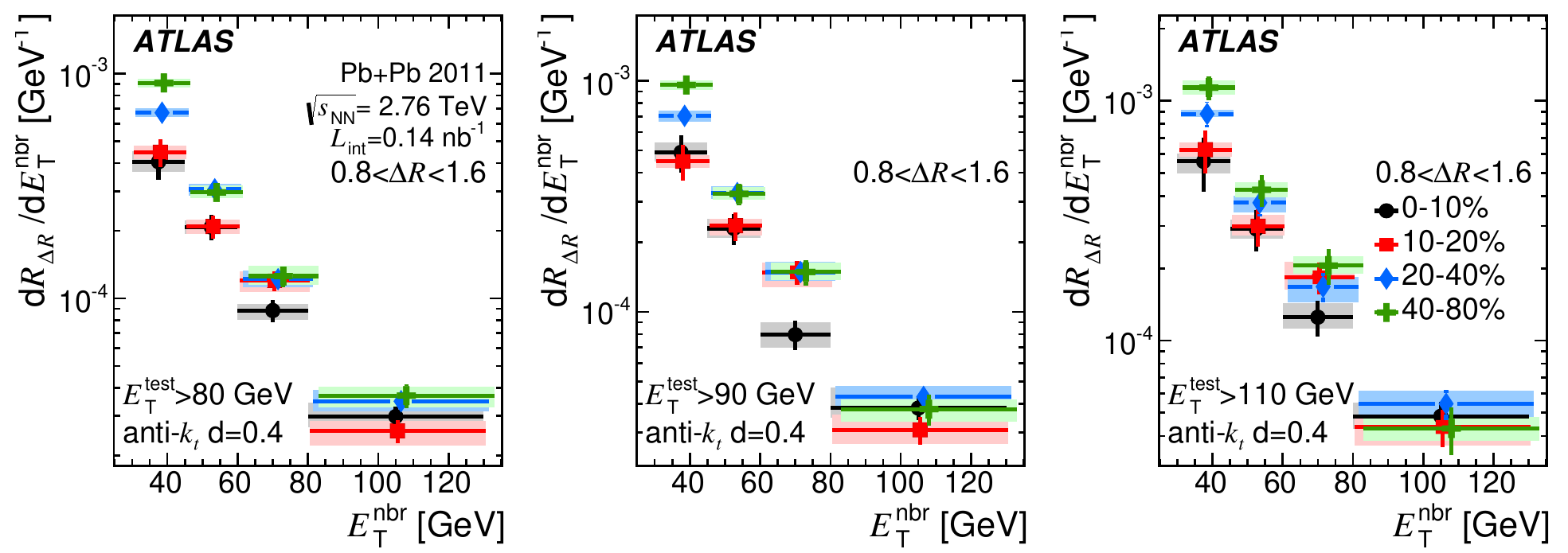}
}
\centerline{
\includegraphics[width=0.8\textwidth]{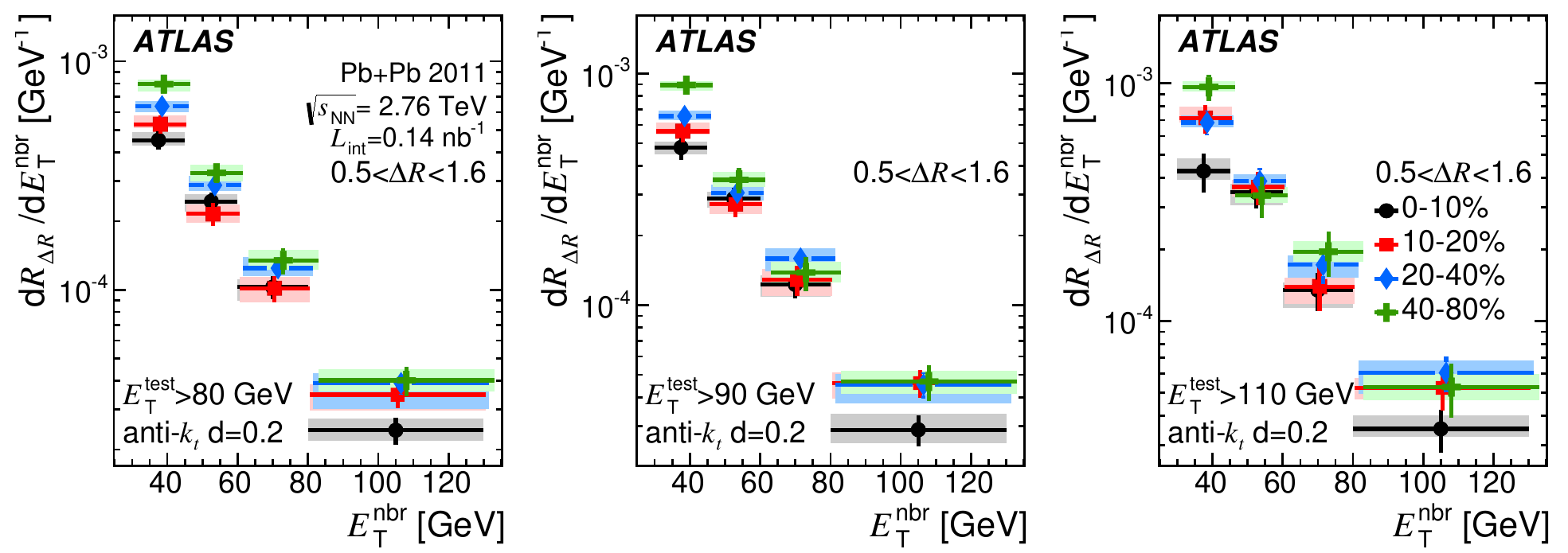}
}
\vspace{-0.15in}
\caption{ 
The  $\mathrm{d} \RdR/\mathrm{d} \ETnbj$ distributions for \RFour\ jets (upper) and \RTwo\ jets (lower) evaluated as a function of 
\ETnbj . The three different columns show the $\mathrm{d} \RdR/\mathrm{d} \ETnbj$ distributions evaluated for three different lower bounds on the test-jet transverse energy, $\ETtest > 80$, 90, and 110~\GeV. The four different centrality bins are denoted by different markers in each plot. The shaded bands 
indicate systematic uncertainties, vertical error bars represent statistical 
uncertainties. 
  The data points and horizontal uncertainties for 10--20\%, 20--40\%, and 40--80\% centrality bins are shifted along the
horizontal axis with respect to 0--10\% centrality bin for clarity.
  }
\label{fig:TJR4R2}
\end{figure*}

To further investigate the centrality dependence of \nearby jet yields, the per-\test-jet normalized \Et\ spectra of \nearby jets defined in Eq.~\ref{eq:2} were evaluated. The resulting differential \Et\ spectra are shown in Fig.~\ref{fig:TJR4R2} for \RFour\ and \RTwo\ jets 
and three different lower bounds on the \test-jet transverse energy, $\ETtest > 80$, 90, and 
110~\GeV. The same trend of suppression in central collisions can be seen as that for \RdR\ evaluated as a function of \test-jet transverse energy shown in Fig.~\ref{fig:OJR4R2}. 
This is a consequence of the steeply falling shape of the \Et\ spectra. To better quantify the differences in 
the \Et\ spectra of \nearby jets, the \Et\ spectra were fitted to a power-law function, 
$\propto 1/\Et^{n}$, and the power index was extracted for all three choices of jet radius and four 
centrality bins. The results are given in Table~\ref{tbl:index}.
  The \Et\ spectra measured in central and peripheral collisions differ in 
the power-law index by approximately two standard deviations for both the \RFour\ 
and \RThree\ jets, suggesting that the \Et\ spectra may be less steep in central collisions than in peripheral collisions. 

\begin{figure*} [ht]
\centerline{
\includegraphics[width=0.8\textwidth]{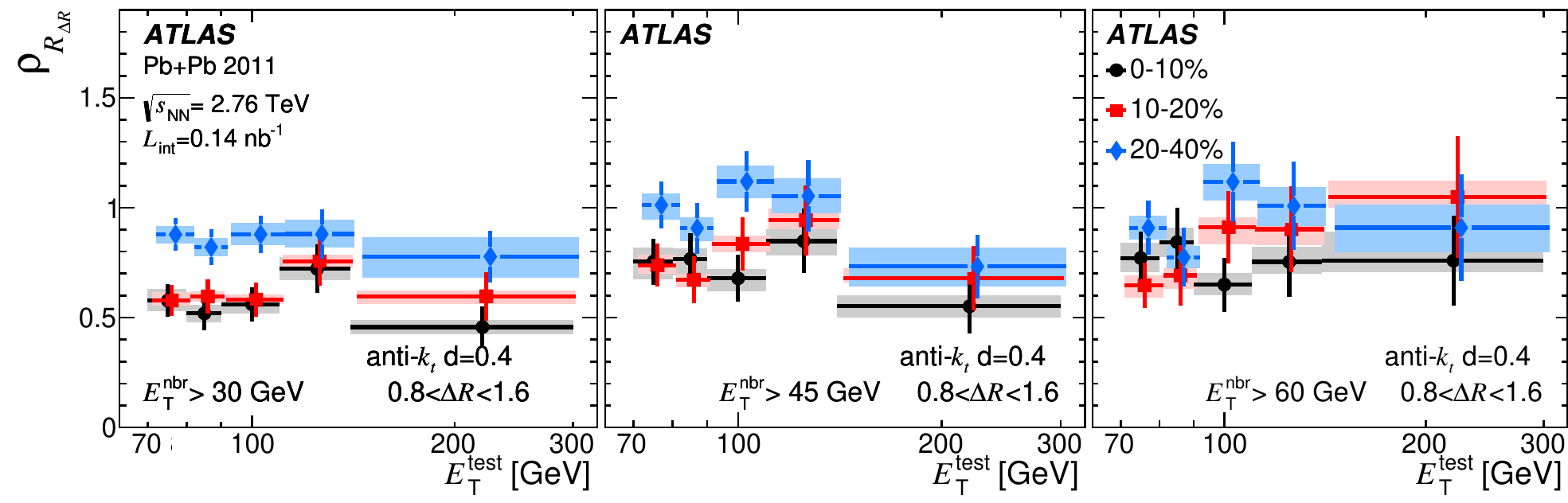}
}
\centerline{
\includegraphics[width=0.8\textwidth]{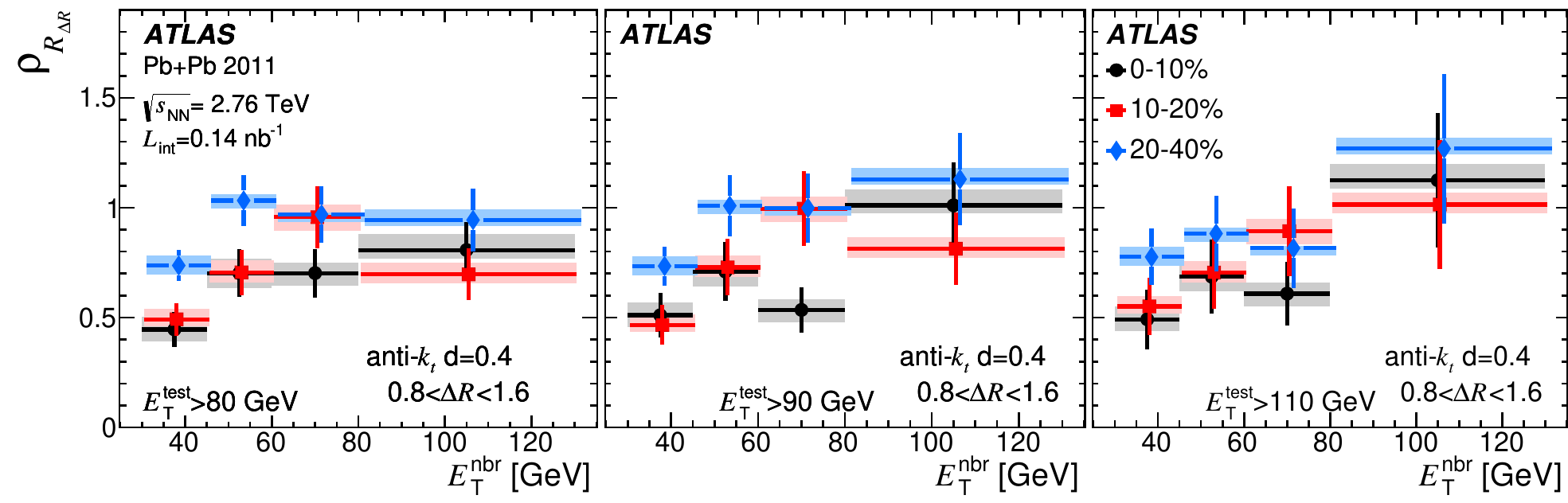}
}
\vspace{-0.15in}
\caption{ 
  The ratio of \RdR\ for three bins of collision centrality to those in 40--80\% collisions, $\RRdR = 
\RdR|_\mathrm{cent}/ \RdR|_{40-80}$ for \RFour\ jets. The \RRdR\ is evaluated as a function of \ETtest\ 
for three different choices of lower bound on \ETnbj\ (upper) and as a function of \ETnbj\ for three 
different choices of lower bound on \ETtest\ (lower). The shaded bands  
indicate systematic uncertainties, vertical error bars represent statistical uncertainties. 
  The data points and horizontal uncertainties for 10--20\%, 20--40\%, and 40--80\% centrality bins are shifted along the horizontal axis with respect to 0--10\% centrality bin for clarity.
  }
\label{fig:rho}
\end{figure*}

To quantify the centrality dependence of the suppression of \nearby jets, the ratios \RRdR\ were calculated by dividing \RdR\ measured in each centrality bin, except the peripheral bin, by \RdR\ measured in the peripheral (40--80\%) bin. 
  Fig.~\ref{fig:rho} shows \RRdR\ evaluated as a function of \ETtest\ and \ETnbj .
  Some systematic uncertainties cancel in the central-to-peripheral ratio as described in 
Section~\ref{sec:systerr}, resulting in \RRdR\ distributions that are dominated by statistical 
uncertainties. Ratios are evaluated for \RFour\ jets, which suffer the least from the statistical uncertainties, that are still large.
  Nevertheless, several characteristic features can be 
observed: the \RRdR\ distributions do not exhibit any strong dependence on \ETtest; the suppression factor 
\RRdR\ of the most central collisions is at the level of 0.5--0.7 for all three lower bounds on \ETnbj; the 
suppression becomes less pronounced with decreasing centrality.
  This is qualitatively consistent with the observation of the centrality-dependent suppression of inclusive 
jet yields \cite{jetsupp}. In that measurement, the suppression of the inclusive jet yields was evaluated 
in terms of the ratio \Rcp\ of the inclusive jet yield in central collisions to the yield in 60--80\% peripheral collisions spanning the jet \pt\ range of 40--200~\GeV. Values of $\Rcp \sim 0.5$ were measured in the 0--10\% most central collisions and exhibited only a weak jet-\pt\ dependence. 

  Contrary to a modest dependence of \RRdR\ on the \test-jet \Et , the \RRdR\ evaluated as a function of 
\ETnbj\ suggests a decrease of suppression with increasing \ETnbj. 
  Such a decrease in suppression with increasing \ETnbj\ may in fact be 
expected. The jet quenching is generally expected to depend on the initial 
parton energy, but if the splitting happens such that the two partons 
have similar energy, their quenching would likely be comparable due to similar 
in-medium path-length travelled by the two partons forming \nearby jets. Thus, 
in the configuration of $\ETnbj \approx \ETtest$ the per-\test-jet 
normalization effectively removes the impact of the suppression.

\section{Conclusions}
\label{sec:conclusions}

  This \papertype\ presents results of a measurement of the production of \nearby jets using pairs of jets produced at opening angles less than $\pi/2$ in $\eta-\phi$ plane. 
  After subtraction of combinatorial backgrounds from different hard-scattering processes, such jet pairs result from the production of multiple jets in the same 
hard-scattering process.
  As such, it is complementary to previous studies of single-jet suppression and dijet asymmetry. By probing the relative 
quenching of a pair of correlated jets in different collision centralities, this measurement opens up the possibility to study the role of fluctuations in the jet quenching process.
This measurement represents a first, exploratory study of how the quark--gluon plasma influences the production and/or later evolution of the \nearby jets from the same parton shower in heavy-ion collisions.

The jet angular correlations were measured in $\sqrtsnn = 2.76$~\TeV\ \PbPb\ 
collisions using $0.14~\invnb$ of data recorded in 2011 by the ATLAS detector at the LHC. The measurements were performed using jets reconstructed with the \antikt\ algorithm for jet radii \RTwo, \RThree, and \RFour . The production of pairs of correlated jets was quantified using the rate of \nearby jets that 
accompany a \test jet, \RdR, evaluated both as a function of \test-jet \Et\ and 
\nearby-jet \Et. A significant dependence of \RdR\ on collision centrality is 
observed in both cases, suggesting a suppression of \nearby jets which increases with increasing 
centrality of the collision. 
  The centrality dependence of the suppression was further quantified using the central-to-peripheral ratio of 
\RdR\ distributions, \RRdR. 
  The trends seen in \RRdR\ evaluated as a function of \nearby-jet \Et\ indicate a decrease in suppression with 
increasing \nearby-jet \Et\ which is, however, of limited significance due to the limited size of the available data sample. The \RRdR\ evaluated as a function of \test-jet \Et\ exhibits a suppression reaching values 
of $0.5-0.7$ in 0--10\% central collisions and does not show any strong dependence on \Et. This behaviour of the \nearby jet production can be used to constrain the theoretical models aiming to describe fluctuations in the jet energy loss.

\section*{Acknowledgments}

We thank CERN for the very successful operation of the LHC, as well as the
support staff from our institutions without whom ATLAS could not be
operated efficiently.

We acknowledge the support of ANPCyT, Argentina; YerPhI, Armenia; ARC,
Australia; BMWFW and FWF, Austria; ANAS, Azerbaijan; SSTC, Belarus; CNPq and FAPESP,
Brazil; NSERC, NRC and CFI, Canada; CERN; CONICYT, Chile; CAS, MOST and NSFC,
China; COLCIENCIAS, Colombia; MSMT CR, MPO CR and VSC CR, Czech Republic;
DNRF, DNSRC and Lundbeck Foundation, Denmark; EPLANET, ERC and NSRF, European Union;
IN2P3-CNRS, CEA-DSM/IRFU, France; GNSF, Georgia; BMBF, DFG, HGF, MPG and AvH
Foundation, Germany; GSRT and NSRF, Greece; RGC, Hong Kong SAR, China; ISF, MINERVA, GIF, I-CORE and Benoziyo Center, Israel; INFN, Italy; MEXT and JSPS, Japan; CNRST, Morocco; FOM and NWO, Netherlands; BRF and RCN, Norway; MNiSW and NCN, Poland; GRICES and FCT, Portugal; MNE/IFA, Romania; MES of Russia and NRC KI, Russian Federation; JINR; MSTD,
Serbia; MSSR, Slovakia; ARRS and MIZ\v{S}, Slovenia; DST/NRF, South Africa;
MINECO, Spain; SRC and Wallenberg Foundation, Sweden; SER, SNSF and Cantons of
Bern and Geneva, Switzerland; NSC, Taiwan; TAEK, Turkey; STFC, the Royal
Society and Leverhulme Trust, United Kingdom; DOE and NSF, United States of
America.

The crucial computing support from all WLCG partners is acknowledged
gratefully, in particular from CERN and the ATLAS Tier-1 facilities at
TRIUMF (Canada), NDGF (Denmark, Norway, Sweden), CC-IN2P3 (France),
KIT/GridKA (Germany), INFN-CNAF (Italy), NL-T1 (Netherlands), PIC (Spain),
ASGC (Taiwan), RAL (UK) and BNL (USA) and in the Tier-2 facilities
worldwide.

\bibliography{DijetAngCorr}
\bibliographystyle{elsarticle-num}

\onecolumn
\clearpage
\newpage 
\begin{flushleft}
{\Large The ATLAS Collaboration}

\bigskip

G.~Aad$^{\rm 84}$,
B.~Abbott$^{\rm 112}$,
J.~Abdallah$^{\rm 152}$,
S.~Abdel~Khalek$^{\rm 116}$,
O.~Abdinov$^{\rm 11}$,
R.~Aben$^{\rm 106}$,
B.~Abi$^{\rm 113}$,
M.~Abolins$^{\rm 89}$,
O.S.~AbouZeid$^{\rm 159}$,
H.~Abramowicz$^{\rm 154}$
% ...
\bigskip
\\
% ...
$^{11}$ Institute of Physics, Azerbaijan Academy of Sciences, Baku, Azerbaijan\\
% ...
$^{84}$ CPPM, Aix-Marseille Universit{\'e} and CNRS/IN2P3, Marseille, France\\
% ...
$^{89}$ Department of Physics and Astronomy, Michigan State University, East Lansing MI, United States of America\\
% ...
$^{106}$ Nikhef National Institute for Subatomic Physics and University of Amsterdam, Amsterdam, Netherlands\\
% ...
$^{112}$ Homer L. Dodge Department of Physics and Astronomy, University of Oklahoma, Norman OK, United States of America\\
$^{113}$ Department of Physics, Oklahoma State University, Stillwater OK, United States of America\\
% ...
$^{116}$ LAL, Universit{\'e} Paris-Sud and CNRS/IN2P3, Orsay, France\\
% ...
$^{152}$ Institute of Physics, Academia Sinica, Taipei, Taiwan\\
% ...
$^{154}$ Raymond and Beverly Sackler School of Physics and Astronomy, Tel Aviv University, Tel Aviv, Israel\\
% ...
$^{159}$ Department of Physics, University of Toronto, Toronto ON, Canada\\
% ...
\end{flushleft}


\end{document}